# The Einstein's Quantum: How can Mathematics be Used to Articulate Reality?

Nathan Willig Lima¹, Tommaso Venturini², Fernanda Ostermann¹, Claudio José de Holanda Cavalcanti¹

1- Federal University of Rio Grande do Sul

2-Institut National de Recherche en Informatique et en Aautomatique

**Abstract:** In this paper, we discuss the role of Mathematics in articulating reality in theoretical Physics. We propose a parallel between empirical and theoretical work and investigate how scientists can also speak about reality without performing any laboratory trial, a key explanation element of STS. To do so, we examine Einstein's 1905 paper on the nature of light for which he received the Nobel Prize of Physics in 1922 and which is deemed as revolutionary by contemporary textbooks. Using Bakhtin's Philosophy of Language, we analyze Einstein's narrative to trace the mechanisms he has used to articulate a new entity, the *quantum*, without performing any experiment or empirical observation. We dialogue with classical results obtained by STS in the context of empirical sciences, drawing in particular on Bruno Latour's concept of 'chain of reference'. We have also used Bakhtin's metalinguistic analysis to highlight Einstein's rhetoric strategies. Our results indicate that the concept of chain of reference can be applied to theoretical physics and that it is possible to trace a parallel between laboratory trials and mathematical trials, in which mathematic operators play the role of laboratory equipment. We show that the main features of the laboratory trials are also present in the mathematical ones. Moreover, our analysis challenges the common representation of Einstein's paper as a moment of epistemological rupture and highlights, on the contrary, its translation efforts to articulate new ideas with the dominant paradigm of the time.

## 1. Introduction

In the late seventies, the focus of Science and Technology Studies (STS) shifted (or rather was extended) from the investigation of scientific knowledge (Bloor 1982 and 1991, Douglas 1970, Cardwel 1971, Shapin and Schaffer 1985) to the observation of scientific practice through ethnographic techniques (Knorr-Cetina 1981, 1995, 1999, Latour and Woolgar 1988, Galisson 1997, Collins and Evans 2002). By focusing on the "anthropology of the laboratory", this "second wave of STS" (Mody, 2015) privileged the description of experimental sciences in detriment of more formal or theoretical disciplines (which rely more on mathematics than on laboratory trials)[1].

To be sure, this does not mean that these disciplines have been completely neglected by STS. There are several studies which discuss the sociological aspects of mathematical proofs (Bloor 1991, 1973, 1978, Livingston 1986, MacKenzie 1999); propose ethnographies of theoretical physics (Gale and Pinnick 1997, Cetina and Merz 1997, Merz and Cetina 1997, Pickering 1999); describe the way in which tools are used (Kaiser, Ito, and Hall 2004) and investigate the relation between physics and its culture (Reyes-Galindo 2014). Few of them, however, tried to extend to theoretical sciences the ontological debate that STS opened on empirical sciences and to explore how scientific facts can be constructed other than by the experimental devices of the laboratory. In this paper, we are specifically interested in reflecting on the following question: how is it possible for a theoretical scientist to talk about the world without performing any experiment? In other words, how can mathematics be used instead of laboratories to articulate reality?

---

[1] This is recognized by Mackenzie (1999) and by Latour (1988) for instance.

We intend to make a small contribution to STS by suggesting a preliminary answer to this question. We will do so by carrying out a metalinguistic analysis (Bakhtin 2016, 2017) of the paper in which Albert Einstein (1905a) proposed that light is composed by "quanta" and for which he received the Nobel Prize of Physics. In our analysis, we discuss the role of mathematics in the articulation of a new actant (the quantum) departing from the concept of "chain of reference", proposed by Bruno Latour (1999)[2].

The choice for Einstein's paper as an object of study is due to its importance in the field of quantum physics according to textbooks (Eisberg and Resnick 1985, Cohen-Tannoudji, Diu, and Laloë 1991, Messiah 1961) and historiographic books (Greenstein and Zajonc 1997, Martins and Rosa 2014). Quantum physics is one of the most popular branches of modern physics, offering what physicists claim to be the most complete theory available to explain matter structure. Through the 20th century, its development originated new areas of physics such as quantum field theory (Landau and Lifchitz 1966, Sakurai 2013), quantum optics (Glauber 1963 a, b) quantum information theory (Benatti 2009), quantum thermodynamics (Vinjanampathy and Anders 2016), quantum gravitation (Woodard 2009), as well as many technological applications in areas like nuclear engineering, semiconductors physics, medicine (Young 1984) and nanoscience (Hornyak, Dutta, and Tibbals 2008). By analyzing Einstein's work, we intend to interpret one of the basilar papers of modern physics.

In section 2, we introduce Latour's concept of "circulating reference" and present key elements of Bakhtin's metalinguistics which we use to interpret Einstein's paper in section 3. In section 3, we dissect Einstein's proposition on the light quanta and we discuss how mathematics can be used to articulate reality. In section 4, we use Bakhtin's philosophy of language to analyze the rhetoric strategies used by Einstein to convince his peers of the quantum existences. In section 5, we' discuss the role of Einstein's paper in Quantum Physics History and its relation to contemporary scientific textbooks.

## 2. Reality in STS research and Bakhtin's metalinguistic interpretation

In order to dialogue with STS debate on ontology, we take as our point of departure the notion of "chain of reference", proposed by Latour after following a soil science expedition in Brazil (Latour 1999). Through its ethnography, Latour concluded that scientific knowledge is not acquired by direct observation, but by subjecting natural phenomena to a series of transformations, which make observation increasingly indirect. It's only through this movement that scientific knowledge becomes possible. Thus, in order to "know the forest", it is necessary to take some distance from it, in a scientific version of Saramago's quote: "you have to leave the island in order to see the island."

---

[2] We could have used any other concept proposed by STS along the ontological discussions in the context of empirical sciences to perform a parallel with the theoretical sciences. The choice for Latour's chain of reference is due to the fact that it is a traditional study in STS, which makes the dialogue with the literature easier.

Never, not even at the beginning of their research, scientists did perform an unmediated observation of their object. From the onset, several maps mediated their appraisal of the Brazilian forest. Their subsequent steps were to collect and arrange samples of the forest (rocks, land and plants) using different tools. At each stage, scientists referred simultaneously to the samples, the code, the instruments, the cartographic coordinates and so on, working on objects that were always hybrids of matter and form, object and sign. At each stage, the obtained hybrid was a sign of the previous step, working as "rough matter" for the next. At each step, scientists discard information about the particularity, materiality and locality of the forest, but they also amplify the compatibility, transportability and generalization of their results.

Though it is not possible to establish a one-to-one correspondence between the different stages of this process (because each implies information suppression, information gain, material transformation, spatial dislocation, etc.), it remains possible to retrace the chain of transformations and reverse it. It is this reversibility that determines the soundness of scientific results.

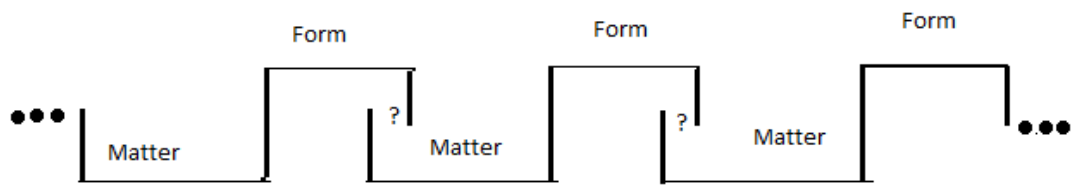

**Figure 1**. Chain of circulating reference. In each step, a new hybrid of matter and form is produced. Between each step and the next, there is an irreconcilable gap. The chain however is reversible, and it is possible to follow all the steps backwards. The figure is adapted from Latour (1999).

The longer the chain is, the farther we are from the "rough matter"; the more we know about something, and thus, paradoxically, the closer we are to it. One of these steps of this chain, however, is distinctively different from the others because it produces the transformation of the samples into a series of signs on a piece of paper (and, more recently, in a digital file). These "inscriptions", as Latour calls them (Latour and Woolgar 1988) can take the form of texts, lists, diagrams, graphics, equations, and many others. What counts is that they are much easier to transport and manipulate than the original samples, and thus replace them completely. Crucially, inscriptions are also codified in ways to facilitate their further transformation through textual or mathematical operations. As all researchers know, the work in the laboratory or in the field does not conclude the research efforts. Once the "empirical" operations are completed (and even before), there is still a great deal of polishing and formatting to be carried out through writing and calculation before findings can be published. While most STS studies have focused on the transformations that lead to the production of inscriptions, in this article we are most interested to the part of the reference chain that follows their creation.

How is it possible to keep transforming signs on papers and still produce knowledge about the "world"? This is precisely what Einstein seems to be doing. He starts in a world with continuous wave light and he ends with a world of quanta, without ever entering a laboratory. But is he still following the same reversible chain of transformation involving gains and losses of information as the previous empirical steps? Or does he develop a different sort of science, more rational and formal as common sense would expect from mathematical reasoning (MacKenzie 1999)? In which ways mathematical tools are similar or different from the experimental devices of the reference chain?

## 2.1. Metalinguistic Interpretation

In order to answer these questions, we perform what Russian semiotician and philosopher of language Mikhail Bakhtin calls a metalinguistic interpretation (Bakhtin 2016, 2017): the analysis of the discursive characteristics of a text, such as theme, organization and style in order to highlight not only its internal structure but also how it affects the external world. Although Bakhtin has focused his analysis on the relation between verbal communication and human culture, we propose to expand his philosophy to the relation between science and society. Through our metalinguistic interpretation, we will try to show how Einstein's text and its mathematics bring a new actant (the quantum) into existence.

Crucial to our argument is the notion of "exotopy': the personality of the "hero" of a text is determined not only by the culture in which he lives but also by the author's "excess of vision". Every text, therefore, tells its readers something not only about its hero but also about the author's vision. In section 3, we carry out an analysis of the way Einstein constructs his hero (the quantum) to reveal his vision of physics. Furthermore, texts are never produced in a "discursive vacuum", but are always part of a dialogue and directed to a specific audience (Bakhtin 2016). This discursive context affects the lexical, phraseological and syntactical choices made by the author. In section 4, we analyze the stylistic elements of Einstein's paper and describe how they are used to uphold the quantum existence.

## 3. Einstein's 1905 paper: Concerning a Heuristic Point of View Toward the Emission and Transformation of Light

The controversy about the nature of light in modern science can be traced back to Newton's and Huygens debate in the XVII century. While Newton used a particle-like model to explain geometrical optics, Huygens explained refraction and diffraction using the concept of "wavefront" (the set of points identically affected by a wave at a given moment). Huygen's vision was corroborated in the XIX century, when Fresnel offered empirical and theoretical arguments in favor of the wavelike model and when James Clark Maxwell has proposed his electromagnetic theory and his notion of "continuous electric and magnetic fields".

Through Maxwell's equations, it is possible to show that both electric and magnetic fields satisfy a wave equation in empty space. More specifically, the electromagnetic wave predicted by this mathematical manipulation has an approximant velocity of $3.10^8$ m/s, the measured speed of light, which for Physics community represented strong evidence that the light was an electromagnetic wave. When Einstein published his paper in 1905, Newton's particle-like model of light had been long deserted.

On the other hand, at the end of XIX and beginning of XX century, thermostatistics was flourishing. According to this approach, matter is made of particles (such as atoms and electrons) with properties (like position and velocity) whose mean values could be connected to macroscopic properties like temperature, pressure and so on. When electromagnetic theory and thermostatistics were used together, however, the theoretical prediction about the behavior of the *black body radiation* not only disagreed with the empirical existing data but also conduced to unacceptable physical results such as predicting an infinite energy in a small volume (the so-called "ultraviolet catastrophe").

In 1905, three models existed to describe black body radiation (Kuhn 1978). The first one was Rayleigh's model, which was based on electromagnetic and thermodynamics theory and that leaded to the mentioned absurd results. The second one was Wien's model, which fitted high frequency radiation (ultraviolet light) data well but failed on low frequency radiation. The third model, developed by Max Planck in 1900 and 1901, fitted all the frequencies with a great accuracy, but its physical interpretation was unclear.

In the paper we are about to analyze, Einstein makes two very counter-intuitive and counter-inductive moves[3]. First, he proposes a sort of Newtonian particle-like conception of light and, second, he rejects Planck's model in favor of Wien's, despite its worse results for low frequency radiation. Yet, without leaving his bedroom, Einstein mobilizes his quanta so effectively to impose his counter-intuitive conception. In the next pages, we will follow each step of Einstein narrative in order to understand how such achievement was possible.

### 3.1. Presentation of a metaphysical perspective

Einstein begins the 1905 paper by posing a metaphysical issue. His first considerations are not about empirical data or mathematical formalism, but about the way in which Physics described reality at the time:

> A profound formal distinction exists between the theoretical concepts which physicists have formed regarding gases and other ponderable bodies and the Maxwellian theory of electromagnetic processes in so–called empty space (…) According to Maxwellian theory, energy is to be considered a continuous spatial function in the case of all purely electromagnetic phenomena including light, while the energy of a ponderable object should, according to the present conceptions of physicists, be represented as a sum carried over the atoms and electrons (EINSTEIN, 1905a, p.132).

While having two kinds of entities in the world (continuous energy and discrete entities) does not imply any logical violation, this distinction is the main feature to be avoided. To Einstein, Physics should provide a unified description of reality (Isaacson 2007), a metaphysical perspective that can be traced back to Parmenides's philosophy (Feyerabend 2010). In Einstein's most important papers[4], this sense of unification seems to be the key element. It is exclusively from this perspective that Einstein decides what is "problematic" and what is not, which direction must be followed, and which must be given up. This can be interpreted as what Mikhail Bakhtin calls "excess of vision" (Bakhtin 1997), i.e. the fact that the literal identity of the protagonist of a text (the light in our example) is determined by elements that are independent from the hero himself, but dependent on the author (who fills the gaps and that translates what he observes). As we will show, the existence of quantum is not something to be demonstrated but it is the 'principle' that guides Einstein's narrative. In order to guarantee the unification of Physics, Einstein introduces the following proposition:

> In accordance with the assumption to be considered here, the energy of a light ray spreading out from a point source is not continuously distributed over an increasing space but **consists of a finite number** of energy quanta which are localized at points in space, which move without dividing, and which can only be produced and absorbed as complete units. (EINSTEIN 1905a, p.133)

---

[3] Counter induction is a concept introduced by Feyerabend (2011) to explain how science can evolve. Instead of evolving through the increasing of empirical consistency as it would argue Popper and Lakatos, to Feyerabend science 'revolutions' take place despite of or even against empirical data. In this sense, science sometimes is counter inductive.

By proposing to consider *a priori* the light as a finite number of energy quanta, Einstein reunifies the two types of entities (bodies and energy), and little matters to him that this conception cannot be conciliated with all observed optical phenomena[5]. Metaphysical and aesthetical insights (his excess of vision) seemed more important to Einstein than empirical data, encouraging him to write a text which is not only counter-intuitive, but also "counter-inductive" (Feyerabend, 2011).

If the positivist description of science were right, the success of modern Physics would be grounded in the absence of metaphysics. As the first paragraphs of Einstein's paper suggest, that it not the case for the proposition of quantum. Its existence is motivated by a metaphysical conception and not by any experimental result (to some extent, even despite experimental results). In this sense, Gabriel Tarde (1999) discussion on modern science seems to apply to Einstein's work. According to Tarde, modern science success cannot be attributed to the adoption of a positivist methodology but should be interpreted as consequence of the pursue of a monadological description, i.e., science evolves when it searches for the minimum element of a certain system, its monad (Tarde, 1999). In the 1905 paper, Einstein is proposing the quantum, the "monad of light". Monadology is not what Einstein "finds" but what he assumes from the very beginning.

Einstein also suggested that light wave-like phenomena result from interactions of many quanta (and not a single quantum) and stick to this position without being able to demonstrate it in subsequent papers (Einstein 1906, 1907, 1909a, 1909b). In 1986, Physics community had empirical evidence that Einstein was partially wrong: despite of corroborating the conception that light is made of quanta, Grangier, Roger, and Aspect (1986) showed that one single quantum can present wavelike behavior.

---

[4] In another 1905 paper, this one about the Special Theory of Relativity (Einstein 1905b), Einstein discusses the unification Electromagnetic phenomena description in different inertial references. And his 1916 General Relativity paper (Einstein 1916) deals with the unification of physical description in any referential (Latour 1988)

[5] Optical phenomena (interference, refraction, diffraction) were considered a strong corroboration of continuous light theory. As we will discuss in section 4, Einstein was not able to provide an alternative explanation for optical phenomena using the concept of *quantum.*

### 3.2. Disarticulation of Continuous Light

To proceed with unification of Physics, Einstein had not only to articulate the quantum, but also to disarticulate continuous light, and, once again, he does not go to the laboratory, but rely on paper and pencil instead. In Section 1, "*Concerning a Difficulty with Regard to the theory of Blackbody Radiation*", Einstein departs from two equations. The first one was obtained in the context of the kinetic theory of gases, and it is the result of a long chain of transformations that are not reported in his paper.

According to Einstein, this equation describes one third of the electron mean energy ($\bar{E}$) as a function of the temperature (T) when the electron is in equilibrium with the molecules of the body:

$$\bar{E} = \left(\frac{R}{N}\right) T \qquad (1)$$

It is important to notice that the signs in this equation do not have a self-evident meaning: they become meaningful only *in relation to* other signs. $\bar{E}$, for instance, can only be expressed in relation to N (the number of molecules in a gram equivalent), R (the universal constant of gases) and T (temperature). Also, the meaning of $\bar{E}$ can be translated by a set of words (which are also signs), "$\bar{E}$ represents one third of the electron mean energy". Each word in the last sentence may be translated into a new set of words such as "electron is an elementary particle with negative charge." Each word of this new sentence may be translated into a new set of words, and so on. In the same way in which the referent of a scientific fact is what circulates through the transformation chain, the referent of a sign is what circulates through a chain of signs (Eco 1991, 30). Thus, after the inscription, the chain of reference keeps moving along, but now through a semiotic chain. Each new sign translates the signs previously constructed, but it is not identical to them. However, that is not enough to explain how Einstein provides new information about the world. To do so, we must observe Einstein's subsequent steps.

The second equation that Einstein introduces (2) was derived by Planck for dynamic equilibrium between electron oscillators and radiation:

$$\bar{E}_\nu = \left(\frac{L^3}{8\pi\nu^2}\right) \rho_\nu \qquad (2)^6$$

Again, Einstein does not explain how Planck obtained this equation: he just informs the readers of its existence. The two equations presented above have been obtained through two different chain of transformations and, so far, are not related to each other. They are totally independent and have no association to each other. What Einstein does next is exactly to connected both unrelated references chains. To do so, he states *aprioristically* that black bodies have their electrons in equilibrium both with molecules and with radiation. Again, this is not something that can be recognized from experiments, it is an *a priori* assumption, what Bakhtin would call *excess of vision*. Supposing that, he can assume that the electron energy can be described by both equations and that (1) and (2) must be equal:

$$\bar{E}_\nu = \left(\frac{L^3}{8\pi\nu^2}\right)\rho_\nu = \bar{E} = \left(\frac{R}{N}\right)T \qquad (3)$$

---

[6] Where $\bar{E}_\nu$ is the average energy of an oscillator with eigen frequency ν, $L$ is the velocity of light, $\nu$ the frequency of the oscillator, $\rho_\nu$dv the energy per unit volume of that portion of the radiation with frequency between ν and ν + dv.

Two separate transformation chains that had no relation to each other are now connected in one single expression that cannot be obtained by experimental processes. This connection allows to describe the relation between energy density and temperature (equation 4), which was not possible when both chains were considered to be unrelated. This gives a first insight on how theoretical physicists can talk about the world without mobilizing experimental equipment: they connect different transformation chains through the translation of an excess of vision. If knowledge comes from the movement through a chain, theoretical physics allows us to know more about the world by moving in a network of transformation chains.

Besides their "extrinsic meaning" (which defines their sense in relation to other hybrids in the reference chain), physics equations also have a different semiotic dimension: the mathematical one, which we will call "intrinsic meaning". $\rho_\nu$, for example, has an extrinsic meaning (density of energy), but can also be treated as a mathematical function (intrinsic meaning) and isolated through a trivial manipulation of equation (3):

$$\rho_\nu = \frac{\left(\frac{R}{N}\right)8\pi\nu^2 T}{L^3} \qquad (4)$$

Thus, we have an equation that relates the density function with the frequency of the radiation. While equation (3) emerged from an extrinsic consideration that allowed us to connect different transformation chains, equation (4) derives only from a mathematical manipulation.

Mathematics, however, has more to offer than tautology. While empirical scientists transform hybrids by means of material tools in laboratory trials, theoretical scientists transform mathematical functions through "mathematical operators", i.e. mathematical objects capable of transforming one another, in mathematical trials.

In order to understand Einstein's trial, it is necessary to be acquainted with two operators defined in differential calculus: the derivative (defining the slope of the curve representing function f(x)) and the integral (the area beneath the curve of function f(x) in a given interval). Derivative and Integral are opposite operators. So, we can take a function f(x) and start to transform it by integration. At each new integral, a new function is generated. If, however, we want to trace the mathematical transformation backwards, we can derivate each function to obtain the previous one. We thus create a chain of functions, each one different from the others, through a reversible process performed exclusively on their intrinsic mathematical meaning. On the other hand, when we transform a function by integrating it, we also create a new function with a new extrinsic meaning. For example, knowing the position of an object as a function of time r(t) it is possible to compute the derivative of r(t):

$$v(t) = \frac{dr(t)}{dt}$$

The new function represents intrinsically the rate of variation of the function in relation to the variable *t*, but extrinsically $v(t)$ represents the variation of the position through time, that is to say its velocity. So, $v(t)$ intrinsically is a function, but extrinsically is a new physical property: "velocity". This means that along this chain we are producing new inscriptions without having to do any experiment. This possibility describes a second way in which theoretical physicists can talk about the world without entering the laboratory. By exploring the intrinsic meaning of mathematical symbols, using mathematical operators, they create new functions whose meaning can be then interpreted extrinsically.

In this way, the integral and the derivative operators work as equipment in a laboratory. As this equipment they are hybrids[7] of matter and form working in reversible reference chains. Einstein uses these mathematical operators to submit the density energy function to a transformation by using the integral operator thus obtaining a new hybrid – the energy per unit volume of the radiation:

$$\int_0^\infty \rho_\nu \, d\nu$$
$$= \text{energy per unit volume of the radiation of frequencies going from zero to infinity} \quad (5)$$

Then he substitute equation (4) with equation (5), to compute the integral over all possible values of frequency $\nu$. This integral does not have a finite value though (it diverges):

$$\int_0^\infty \frac{\left(\frac{R}{N}\right) T 8\pi\nu^2}{L^3} d\nu = \infty \quad (6)$$

A diverging integral is not a problem in mathematical terms, but it is in physical terms. The integral "is saying" that the density of energy is becoming infinite, which from the point of view of physics is unacceptable: a body cannot have infinite energy, as this would violate the physical postulate of energy conservation. After all the mathematical manipulations, the final result of the transformation chain is unacceptable.

Mathematical signs, thus, behave in theoretical Physics seemingly to the non-humans that the empirical scientist deal with. According to Latour, for instance, the best evidence that microbes have agency is the fact that they resist to Pasteur's intentions, they do not always act as he expects and wishes (Latour, 1993, 1999, 2005, 2013)[8]. In a similar way, equations resist to theoretical physicists and act independently from their will. Starting from equation (2) and (3), classical physicists expect and wish to arrive at a finite density of energy. Yet, as Einstein pointed out, equations produce an infinite density of energy instead and in spite of classical physicists' desires. What Einstein does is then to exploit this resistance to enroll equations as allies in his disarticulation of the continuous light.

---

[7] Mathematical operators, as every symbol, are also hybrids since all semiotic construction cannot be departed from matter (Bakhtin 2006, Vygotsky 2015, Wertsch 1985, Eco 1991, Latour 1999).

Having highlighted the undesired result obtained by the mediation of classical theory equations, Einstein ought to point which step of the reference chain is invalid. It can be any step. The problem could have come, for instance, from the fact that black bodies are not in thermodynamic equilibrium and thus the equality used in equation 4 is not possible, or maybe that the kinetic theory of gases is wrong (which makes equation 1 wrong). Einstein however assumes that the problem is that we are using the idea of continuous radiation that led Planck to equation (2). This assumption however remains undemonstrated. If continuous light is blamed for the unacceptable final result, it is only because of Einstein's "excess of vision" (his pursuit for unification).

The analysis of the first part of Einstein's paper already provides a preliminary answer to our questions: "how theoretical physicists can speak about the world without entering the laboratory?" We have shown that Einstein uses mathematics to connect different transformation chains, allowing scientific reference to circulate through their network. Furthermore, he explores the two dimensions of physical equations: he transforms equations by using mathematical operators (exploring their intrinsic meaning), and then he has interpreted the result from the viewpoint of physics (exploring their extrinsic meaning). Much like an experimental scientist, Einstein mobilizes a series of hybrid devices (the mathematical operators that constitute his equipment) to produce a chain of reversible transformations. In the end, the result obtained cannot be attributed exclusively to the will of the scientist, because it would not have been possible without the active role of equations.

Through these operations, Einstein has done half of his work: he destabilized the structures of the continuous light model. Now he has another task: to build up the quantum.

### 3.3. Articulating the quantum

What does it mean to articulate the quantum? Again, the parallel between empirical and theoretical sciences holds. When empirical scientists articulate a new entity in their laboratory, they do so by assessing their performance in a series of tests. For instance, when Pasteur articulated microbes, he performed a series of laboratory trials to determine microbe performances. The essence of the microbe in the end is the label that glues all these performances (Latour, 1999). Einstein must do the same to articulate the quantum. To determine the performance of quanta, he must submit them to a series of mathematical trials and then sum up the results through his excess of vision.

---

[8] The conception of non-human agency is far from being a consensual point in Sociology of Science. Besides of Latour (1993, 1999, 2005, 2013), many studies can be found about the non-human agency debate (Bowden, 2015; Jansen, 2016; Dürbeck, Schaumann, Sullivan, 2016).

Einstein's strategy involves showing that high frequency electromagnetic radiation performs like an ideal gas (which is composed by particles). First, he determines the electromagnetic radiation entropy using thermodynamic and Wien's equation, which results in equation 7.

$$(S - S_0)_{radiation} = \frac{E}{\beta \nu} \ln\left(\frac{V}{V_0}\right) \tag{7}$$

Equation 7 states that, for high frequency electromagnetic radiation, entropy behaves as a logarithm function of volume. Also, Einstein shows in his paper how to compute the entropy for an ideal gas, obtaining equation 8:

$$(S - S_0)_{gas} = \frac{nR}{N} \ln\left(\frac{V}{V_0}\right) \tag{8}$$

Equation 8 states that, for an ideal gas, entropy behaves as a logarithm function of volume. Though Einstein has obtained them through very different mathematical operations, they look very similar. Both have the same structure: a constant multiplied by the logarithm of the rate of volumes. From this, all we can say from combining intrinsic and extrinsic meanings is that radiation entropy behaves similarly to gas entropy in relation to volume variation. But Einstein has a deeper metaphysical reading for this result: for him the similarity implies that light consists of quanta in the same way as gas consists of particles. To Einstein, equation (7) and (8) are expressing the same kind of system, which means that both equations should be identical. Thus, by equating them ($\frac{E}{\beta \nu} \ln\left(\frac{V}{V_0}\right) = \frac{nR}{N} \ln\left(\frac{V}{V_0}\right)$), he also assumes that the constant before the logarithm must also equal, which allows him to isolate E as:

$$E = n \frac{R\beta}{N} \nu \tag{9}$$

Einstein has just stated a new performance for the quantum. By connecting the results of two chain of transformation through his excess of vision, he concludes that the energy (E) of the quantum should behave as function of its frequency ($\nu$). This performance, despite of being obtained by manipulating classical theory equations, is a novelty in Physics. Notice that Einstein has not demonstrated that "quantum exists" he has supposed its existence and then uses a series of mathematical trails to determine the performances of this entity. These mathematical trials could not be defined as a rigorous mathematical demonstration, but a creative process in which intrinsic and extrinsic dimensions of mathematical signs are combined. It is necessary now to observe whether the network articulated in this way allows a coherent movement through other inscriptions and chains. That is what Einstein does when he takes phenomena to testify.

Having determined the performances of his quanta, Einstein must show that they are consistent with phenomena related to light emission and transformation. He does so by mobilizing the inscriptions from three phenomena. So, while never setting foot in a laboratory, Einstein draws on the empirical results obtained by other scientists to give his quantum more substance.

Einstein realized that three phenomena related to light emission and transformation were dependent on the frequency radiation – a dependency that could not be explained by classical theories. By articulating an equation that associates energy and frequency (equation 9), Einstein was able to explain them in terms of energy conservation (a very familiar concept for physicists).

First, Einstein explained the so-called Stoke's Rule, an empirical statement according to which when a radiation of frequency $v_1$ is projected on a photoluminescent solid (a solid that starts shining after being irradiated), the frequency of the radiation emitted by the solid ($v_2$) is always smaller than $v_1$. While empirical physicists could only determine that $v_2 < v_1$ but not explaining why, Einstein could explain the this relation by mathematically transforming it in $E_2 < E_1$[9] (the energy of the radiation emitted by the photoluminescent solid is always smaller than the energy of the radiation to which it was exposed), which can easily be understood in terms of energy conservation.

The second phenomenon explained was the Photoelectric Effect, the ejection of electrons by a surface exposed to an electromagnetic radiation. According to the classical electromagnetic theory, electromagnetic waves transport an amount of energy proportional to the amplitude of the electromagnetic field. In contrast with this classic expectation, Einstein's articulation assumed electromagnetic radiation to be composed of quanta with energy proportional to their frequency (and not to amplitude). By assuming this, Einstein was able to predict the relation between the kinetic energy of ejected electrons and the frequency of incident radiation. Milikan confirmed the validity of Einstein's prediction in 1921, and Einstein received the Nobel Prize for it in 1922.

The third phenomenon that Einstein explained was the ionization of gases. Einstein predicted the relation between the number of molecules ionized by a gas and the quantity of absorbed radiation using equation 9. By doing so, he was able to propose another empirical verification of quantum performance.

This concludes the analysis of Einstein's articulation of the quantum. It is interesting to observe that his argument begins with inscriptions concerning energy and radiation. He submits them to a series of mathematical transformations to articulate the performances of an actor (the quantum), and finishes it by using the features of this actor to interpret other inscriptions (three phenomena we just discussed). This corroborates the idea that the theoretical work deals with intrinsic and extrinsic manipulation of mathematical inscriptions to connect different chains of transformation in a single network. In the next section, we highlight rhetoric strategies used by Einstein to strengthen his mathematical construction.

## 4. Rhetoric tactics

Though the mathematical treatment to which Einstein submit his quanta is brilliant, it is not this treatment that constitutes the main contribution of his paper. The equations obtained by Einstein regarding radiation and ideal gas only indicate that both behave similarly in relation to entropy variation. It only through his *excess of vision* that he can connect the two equations and derive an original performance of his quantum. The problem is that this extra-meaning attributed by Einstein to the equation contradicts several key elements of the widely accepted Maxwell's theory of electromagnetic field. If Einstein proposition goes farther than mathematics can go, how can Einstein persuade his peers?

---

[9] In order to achieve this new equality you can multiply $v_2 < v_1$ by $n\frac{R\beta}{N}$ in both sides and use equation 9.

In this section, we discuss the rhetoric tactics employed by Einstein in his 1905 paper. Following Gusfield's (1976) and Latour & Fabbri's (1977) we searched for textual choices adopted by Einstein to strengthen his assumptions. To do so, we rely Bakhtin's philosophy of language according to which, any piece of communication is the combination of a variety of elements (theme, structure and style) mobilized to provoke an intended response from an addressed audience (Bakhtin, 2016). In Einstein we can highlight at least four different rhetoric strategies.

**a. Assuming weakness**. This first strategy is used as early as the headline of the article, "On a Heuristic Point of View Towards the nature of emission and Transformation of Light". In the title, Einstein present his paper as "a Heuristic Point of View Towards". This expression does not add any substantial information about the theme or about the theoretical perspective adopted by Einstein. *Point of View* expresses a subjective perspective and the word *heuristic* suggests that the argument that is not formal and strict (French, 2009). The expression could have been omitted without altering the meaning of the title substantively: its function seems rather to put the whole paper between "scary quotes". As argued in section 3, the quantum existence is less the result of the mathematical articulation than of a "meta-physical" assumption, and Einstein is fully aware of this. By explicitly assuming this weakness of his work, Einstein is anticipating criticism. He is warning the reader of his paper (the community of German physicists) that he is offering just a conjecture, a heuristic point of view.

**b. Flattering opposite ideas**. In the introductory text of Einstein's paper, we can find the following passage, in which we highlighted some interesting lexical choices:

> The wave theory of light, <u>which operates with continuous spatial functions</u>, **has worked well** in the representation of purely optical phenomena, **and will probably never be replaced by another theory**. It should be kept in mind, however, that the optical observations refer to time averages rather than instantaneous values. **In spite of the complete experimental confirmation of the theory applied to diffraction, reflection, refraction dispersion, etc**., it is still conceivable that the theory of light <u>which operates with continuous spatial functions</u> may lead to contradictions with experience when it is applied to the phenomena of emission and transformation of light.

Einstein may have just said that Electromagnetic Theory refers to time averages, and this can lead to contradiction in the description of isolated phenomena involving emission and transformation of light. Instead he decides to flatter Electromagnetic theory by saying that it "works well and it will probably never be replaced" and that it has "the complete experimental confirmation of the theory applied to diffraction, reflection, refraction dispersion, etc." By doing so, it seems that Einstein is trying to create an auspicious atmosphere to introduce his idea. Instead of being a cold report of a mathematical articulation, Einstein paper is a very careful communication, full of rhetorical subtleties.

**c. Reducing electromagnetic theory to a continuous theory**. From the last passage, it is possible to understand how Einstein presents electromagnetic theory as a continuous spatial function. While it is true that Electromagnetic Theory works with continuous spatial functions, it also works with magnetic monopoles and inverse-square law for the Electromagnetic field, using electric unities (the electrons) and so on. Continuous spatial functions are only some of the many attributes of electromagnetic theory.

**d. Shifting rhetorical frames**. A classic distinction in epistemology opposes an "essentialist" view according to which science describes reality from an "instrumentalist" view according to which sciences merely provide a model of reality. For instance, by taking one position or the other, one can say that science *represents* the solar system as if the Sun was in the center and the Earth was moving around it or that science asserts that the Sun *is* in the center of the system and Earth is moving around it. Considering this, let us look at three parts of Einstein's introductory text:

> "The wave theory of light, which operates with continuous spatial functions, has worked well in the **representation** of purely optical phenomena"
>
> "It seems to me that the observations associated with blackbody radiation, fluorescence, the production of cathode rays by ultraviolet light, and other related phenomena connected with the emission or transformation of light **are more readily understood if one assumes** that the energy of light is discontinuously distributed in space."
>
> "In accordance with the assumption to be considered here, the energy of a light ray spreading out from a point source **is** not continuously distributed over an increasing space but **consists** of a finite number of energy quanta which are localized at points in space, which move without dividing, and which can only be produced and absorbed as complete units."

In the first sentence, Einstein uses an instrumentalist lexical choice to speak about the approach he opposes. In the second, he adopts an instrumentalist framing for his own approach. Finally, in the third sentence, he shifts to an essentialist perspective to present his idea. In these sentences, we can observe two rhetoric strategies. The first is a shift of ontological framing: while Electromagnetic Theory deals only with *representation*, his approach deals with an *essential reality*. Einstein, however, does not make it abruptly: he makes a smooth transition through an intermediary phase in which he uses an instrumentalist perspective for his own idea.

Furthermore, a fifth rhetorical strategy can be found in the following papers published by Einstein to convince a skeptical academic community of the existence of the quanta (Martins and Rosa, 2014). To support his view, Einstein not only connected the quanta to other physical entities and gave them new properties – for instance the possibility of transferring momentum (Einstein 1909b) – but also adopted a new rhetoric strategy in relation to Planck. Even though in the 1905 paper Einstein opposed Planck's theory and preferred Wien's model, he later changed his position and tried to side with Plank. In a 1906 paper, for example, he explained:

> This relationship was developed for a range that corresponds to the range of validity of Wien's radiation formula. **At that time it seemed to me that in a certain respect Planck's theory of radiation constituted a counter part to my work**. New considerations, which are being reported in §1 of this paper, showed me, however, that the theoretical foundation on which Mr.Planck's radiation theory is based differs from the one that would emerge from Maxwell's theory and the theory of electrons, precisely because **Planck's theory makes implicit use of the aforementioned hypothesis of light quanta.** (Einstein 1906)

At the time when Einstein was just beginning his career, Planck was a prestigious theoretical physicist. By associating his proposal to Planck's theory (Einstein, 1906, 1907, 1909a and 1909b), Einstein tried to make his idea more palatable to the academic community, despite the fact that Planck himself made clear that he had not agreed with the quantum assumption and that, as later proved by De Broglie, Einstein's atomistic view of electromagnetic radiation is essentially incompatible with Planck's model (Martins and Rosa 2014).

## 5. The stabilization of the quantum hypothesis in contemporary scientific community

The painstaking analysis to which we have submitted Einstein's paper may seem exceedingly nitpicking. After all, any physicist who would take the time to read Einstein's text attentively could observe the points that we have discussed so far. But this is precisely the point: most physicists do not learn the bases of their discipline through the original articles that introduced them, but through the textbooks employed in their academic education (Kuhn, 1996). For sake of clarity, these textbooks tend to present the history of science in any simplified way, neglecting most of the complexities of the original arguments on purpose. For example, introducing Einstein's theory of light, the manual by Eisberg and Resnick (1985) affirms:

> In each case we shall obtain experimental evidence that radiation is particle-like in its interaction with matter, as distinguished from the wavelike nature of radiation when it propagates (Eisberg and Resnick, 1985: 27)

That is exactly the opposite of Einstein's idea, whose main assumption is that light consists *and propagates* in quanta. The view expressed by the authors resemble rather to Planck's original view mixed with the more recent "complementarity principle" (Bohr 1928) according to which particle-like and wave-like descriptions are complementary (an interpretation also defended by prestigious physicists such as Richard Feynman, 2008)

This illustrates how textbooks do not explain Einstein's work by using its original articulation but by combining multiple further interpretations of quantum mechanics (in a sort of "reverse causation", Latour 1999). Textbooks do not rely exclusively on the Einstein's original perspective, neither on any pure contemporary perspective. They hybridize their speech (Bakhtin, 1981) through a composition of different interpretations defended along the decades. Textbooks in this sense contribute to the stabilization of a scientific fact by defusing the theoretical controversies. Instead of making explicit that there are different interpretations for quantum performance and stressing the ontological and epistemological divergences among them, they combine these perspectives, attributing them to Einstein in a way that is glorifying, but historically inexact. Textbooks also tend to present Einstein's theory of light as a revolution:

> at the beginning of the twentieth century, physics was to be marked by the profound upheaval that led to the introduction of relativistic mechanic and quantum mechanics. The relativistic "**revolution**" and the quantum "**revolution**" were, to a large extent, independent since they challenged classical physics on different points (Cohen-Tannoudji, Diu, and Laloë 1991, 9)

Most science manuals present Einstein's paper as a radical innovation compared to classical physics. Is such view supported by Einstein original paper? In his famous book on *Scientific Revolutions*, Thomas Kuhn (Kuhn 1978b) distinguish two dynamics evolution of science: "normal science" and "scientific revolutions". In normal science, scientists struggle to solve specific questions using the conceptual and methodological tools available in the dominant paradigm. Some of questions, however, remain unsolved despite of all the efforts of the scientific community and their accumulation end up generates a crisis and opening a scientific revolutions. During such revolution a new paradigm is introduced (usually through much turbulence) to re-interpret previous empirical data. In Kuhn's words:

> Normal science ultimately leads only to the recognition of anomalies and to crises. And these are terminated, not by deliberation and interpretation, but by a relatively sudden and unstructured event like the gestalt switch. Scientists then often speak of the "scales falling into the eyes" or of the "lightning flash" that "inundates" a previously obscure puzzle, enabling its components to be seen in new way that for the first time permits its solution (Kuhn 1996: 122).

Contemporary scientific textbooks tend to present the blackbody radiation, photoelectric effect, fluorescence and gas ionization as the anomalies that brought about the crisis of the classical physics and Einstein's 1905 paper as the "gestalt switch" that introduced a new incommensurable paradigm.

However, as we have shown in section 2, Einstein's construction of the quanta is fully grounded in previous physics (electromagnetic theory, thermodynamics and Boltzmann's principle). Similarly, Einstein's mathematics does not introduce any major novelty. His paper certainly opposes the idea of a continuous electromagnetic field, but this is far from a radical rupture. Quite the contrary it relies on old concepts and on classic mathematical formalism. It is deterministic and realist. It is not even based on Planck's model, but the older one suggested by Wien.

Einstein's "revolution" is not revolutionary at all. His paper resembles much closer to a "translation" process (Callon 1986) than to an epistemological rupture. As we have tried to show by our analysis, Einstein made all great mathematical and rhetorical efforts to reassure his readers and convince them that his proposal contains as much continuity as rupture. Recognizing these efforts is crucial to realize that the ideas proposed by

Einstein did not imposed themselves by the sheer force of evidence, but by the relentless work of articulation that the physicist invested to associate his new ideas with the old paradigm.

6. **Conclusion**

In this work, we examined the paper in which Einstein proposes the idea of light quantum without performing any experimental trial. In line with Tarde's idea of scientific research (1999), we have shown that Einstein paper, far from being deprived of metaphysics, assumes from the very beginning, a monadological standpoint assuming that the existence of quanta is necessary to unify the physical description of reality. We investigated how it was possible for Einstein to bring this new actor on the scene of physics without the tiniest laboratory experience.

We described how Einstein mobilized extrinsic considerations (what we have called his "excess of vision", according to the ideas of Mikhail Bakhtin) to extend the chain of reference through mathematical transformation and to connect different pre-existing experimental chains in a single network. We also found out that Einstein uses mathematical operators in similar way to which empirical scientists use laboratory equipment: through a sort of "mathematical trials", Einstein explores two dimensions of equations, their mathematical dimension (or "intrinsic meaning") and their physical dimension (or "extrinsic meaning"), articulating at the same time new math functions and new hybrids in the reference chain. This implies that the mathematical work in theoretical physics is not a simple sequence of algorithmic steps, but it is full of uncertainty and "excess of vision" just as like the empirical sciences.

Precisely, we have shown that the symmetry between empirical and theoretical sciences holds at least in three senses. First, mathematical operator transforms mathematical functions in the same way as laboratory equipment transform the subsequent inscription of the reference chain. Second, the theoretical scientist must interpret the new hybrid created by the mathematical operator in the same way the empirical scientist has to fulfill the gaps to interpret an experiment. Third, equations can resist to the theoretical scientist in the same way non-human can resist the experimenter in the laboratory. Also, we showed that the excess of vision the theoretical scientist is responsible for connecting independent reference chains into a single network.

We also have shown that Einstein employs different rhetoric strategies to back his assumptions and to make his position more acceptable. Much of these strategies are devoted to creating continuity between his proposal and the established paradigm of classic physics. Unlike what textbooks often affirm, Einstein paper contains as much continuity as rupture, in a way that it is better described as a translation process rather than an epistemological revolution.